\documentclass[aps,prb,twocolumn,psfig,showpacs]{revtex4}
\usepackage{amsfonts}
\usepackage{amsmath}
\usepackage{graphicx}
\usepackage{bm}
\usepackage{color}
\usepackage{amssymb}
\usepackage{ulem}
\usepackage{times}
\usepackage{dcolumn}
\usepackage{cases}
\usepackage{txfonts}

\begin{document}

\title{Theory of Network Contractor Dynamics for Exploring Thermodynamic Properties of Two-dimensional Quantum Lattice Models}
\author{Shi-Ju Ran, Bin Xi, Tao Liu and Gang Su}
\email[Corresponding author. ]{Email: gsu@ucas.ac.cn}
\affiliation{Theoretical Condensed Matter Physics and Computational Materials Physics Laboratory, School of Physics, University of Chinese Academy of Sciences, P. O. Box 4588, Beijing 100049, China}

\begin{abstract}
Based on the tensor network state representation, we develop a nonlinear dynamic theory coined as network contractor dynamics (NCD) to explore the thermodynamic properties of two-dimensional quantum lattice models. By invoking the rank-$1$ decomposition in the multi-linear algebra, the NCD scheme makes the contraction of the tensor network of the partition function be realized through a contraction of a local tensor cluster with vectors on its boundary. An imaginary-time-sweep algorithm for implementation of the NCD method is proposed for practical numerical simulations. We benchmark the NCD scheme on the square Ising model, which shows a great accuracy. Besides, the results on the spin-1/2 Heisenberg antiferromagnet on honeycomb lattice are disclosed in good agreement with the quantum Monte Carlo calculations. The quasi-entanglement entropy $S$, Lyapunov exponent $I^{lya}$ and loop character $I^{loop}$ are introduced within the dynamic scheme, which are found to display the ``nonlocality" near the critical point, and can be applied to determine the thermodynamic phase transitions of both classical and quantum systems.
\end{abstract}

\maketitle

\section{Introduction}

Two-dimensional ($2$D) strongly correlated quantum models have triggered broad interest in last decades as they always exhibit intriguing and exotic properties (e.g. Refs. [\onlinecite{QHE,QSL}]). Accompanied with the boom of quantum information science, some theories \cite{Entangle,Fidelity} were developed to enable us to describe the critical phenomena beyond the traditional paradigm of Landau-Ginzburg and renormalization group and to access the information of phase transitions without acquiring knowledge of order parameters or universality class. For example, the fidelity is shown to be able to determine the ground state phase diagram \cite{Fidelity}, and the scaling law of the entanglement entropy in $2$D is utilized to identify the topological orders \cite{Topo}.

Another issue that is under hot discussion is about the efficient and controllable numerical algorithms for strongly-correlated systems. While reliable analytical methods for such systems are still sparse owing to the complexity of many-body interactions, numerical means play essential roles. Among others, the quantum Monte Carlo (QMC), density matrix renormalization group (DMRG) \cite{DMRG} as well as its variants \cite{FTDMRG,DMRG2D}, tensor network state (TNS) based algorithms \cite{PEPS,TRG1,TRG2,TRG3,TRG4,MERA,LTRG,ODTNS,HOSRG}, and so on, have achieved great success \cite{ReviewDMRG,ReviewTNS}. However, QMC is not applicable to the frustrated spin systems and Hubbard model away from the half-filling because of the ``negative sign" problem, and DMRG is remarkably accurate and efficient in one dimension but has great costs for $2$D systems of large size. Therefore, to develop new theories and efficient algorithms for correlated quantum lattice systems is highly encouraged.

In this paper, we develop a theory of the network contractor dynamics (NCD) that comprises two aspects: three suggested generic in the NCD theory are introduced and shown to be capable of detecting phase transitions, and an efficient and well-controlled algorithm based on the NCD theory for investigating the thermodynamic properties of $2$D quantum lattice models is proposed. It is demonstrated that the NCD theory is flexible and applicable to the models which can be represented in the form of a tensor network \cite{TRG1,TRG4}, e.g. the $2$D Heisenberg and Ising models.

The primary strategy is as follows. First, we represent the density operator at an infinitesimal inverse temperature in the form of tensor product density operator (TPDO), with which the density operator at finite temperatures can be simulated by imaginary time evolution \cite{ODTNS}. Then the partition function $Z$ as well as the thermal averages of observable operators $\langle \hat{O} \rangle$ can be calculated by contracting all shared bonds in the 2D TPDO. In this course, there are two unavoidable difficulties that both involve in an infinite contraction of the TN which cannot be exactly achieved: (a) the evolution procedure will increase exponentially the bond space of the TPDO; (b) The calculations of $Z$ and $\langle \hat{O} \rangle$ at the targeted temperature include infinite contractions of the TN. Our general strategy to deal with these difficulties is to make the infinite contraction by that of a local cluster with proper vectors on its boundary. To be specific, we first obtain the TN of the partition function from the TPDO, and then consider a properly selected cluster of tensors in this TN as one tensor such that it is the only inequivalent tensor in the TN. Such a tensor (denoted by $\mathbf{T}^{cell}$) can be regarded as a nonlinear mapping over the spaces of different bonds. By using the fixed point of such a mapping that is referred to as the \textit{contractor} and denoted as $\{ \mathbf{\tilde{x}} \}$, we manage to simplify the calculations of $Z$ and $\langle \hat{O} \rangle$ as local contractions of $\mathbf{T}^{cell}$'s and $\{ \mathbf{\tilde{x}} \}$. The simplification of calculating truncations is similar. An imaginary-time-sweep algorithm for implementation of the NCD theory is proposed, which is free from the negative sign problem and whose errors can be well controlled. The results calculated by the NCD scheme for the Ising model on square lattice spin-1/2 Heisenberg antiferromagnet (HAF) on honeycomb lattice are nicely compared with the exact results and other methods as well as the QMC simulations, showing the efficiency and accuracy of this method.

An imaginary-time sweep algorithm is proposed for the implementation of the NCD theory. We show that the three parts of the error are well-controlled. Like other algorithms (e.g. Refs. [\onlinecite{TRG1,LTRG,ODTNS}]), the error brought by the Trotter-Suzuki decomposition is controlled by the imaginary time slice and the truncation error controlled by the discarding weight, but there is no quantity to show how far the truncation is from the global optimal one, i.e., to control the error of the truncation (comparing with the global optimal truncation). In the NCD scheme, all three parts are well-controlled, while the last one is controlled by the loop character. We argue that when the loop character decays to zero, the effect of larger loops (which are destroyed) is negligible and the truncation can be considered as globally optimal.

Within the framework of the NCD, the quasi-entanglement entropy $S$ defined by the transfer matrix of the partition function, Lyapunov exponent $I^{lya}$ that quantifies the convergent properties of the introduced non-linear mapping, and the loop character $I^{loop}$ describing the loop-dependence of the tensor networks are introduced to characterize the properties of the TNS. It is found that these three quantities describe the nonlocality of the quantum states, and are able to detect possible thermodynamic phase transitions of both the classical and quantum systems, as manifested by the square Ising model and the spin-1/2 anisotropic honeycomb HAF model.

The paper is organized as follows. In Sec. II, we briefly present the equivalence between the finite temperature density operator and a TN. In Sec. III, the theory of NCD is introduced based on the TN representation. In Sec. IV, a way to increase the cell tensor size in the NCD, the quasi-entanglement entropy and the loop character are proposed. In Sec. V, we suggest the imaginary-time-sweep algorithm for the implementation of the NCD theory. In Sec. VI, we test the accuracy of the NCD scheme on the square Ising model and the spin $1/2$ HAF on honeycomb lattice and show that the three quantities, $S$, $I^{loop}$ and $I^{lya}$ are able to detect possible thermodynamic phase transitions. Finally, a summary is given.

\section{Equivalence between the finite temperature density operator and a tensor network}

In this section, we show the equivalence between the finite temperature density operator and a tensor network. Henceforth, we take the HAF on honeycomb lattice as an example, and the following discussions below can be readily extended to other 2D lattices.

Suppose that the Hamiltonian of a quantum lattice model with nearest neighbor couplings can be written as $\hat{H} = \sum_{\langle ij \rangle }\hat{H}^{ij}$, where $\hat{H}^{ij}$ is the local Hamiltonian of two connected spins at $i$th and $j$th lattice sites. The calculation of the finite temperature density operator $\hat{\rho}$ can be transformed into the contraction of a three-dimensional TN as follows. Let us begin with introducing the local evolution operator $\hat{U}^{ij} = e^{-\tau \hat{H}^{ij}} = \sum_{p_ip_jp_i'p_j'} U_{p_ip_i'p_jp_j'} |p_ip_j \rangle \langle p_i'p_j'|$, where $\tau$ is the infinitesimal imaginary time slice, $|p_i\rangle$ is the local basis of the $i$th spin and $p_i$ denotes the physical bond. By using the Trotter-Suzuki decomposition \cite{Trotter}, we can write the density operator at inverse temperature $\beta$ as $\hat{\rho}(\beta) = [\prod_{\langle ij \rangle } \hat{U}^{ij}]^K$ with $\beta=K\tau$. Making use of the singular value decomposition (SVD), the matrix $U_{p_ip_i'p_jp_j'}$ can be decomposed into $U_{p_ip_i'p_jp_j'} = \sum_{g} U^{L}_{p_ip_i',g} \lambda_g U^R_{p_jp_j',g}$ , where $g$ denotes the geometrical bond that is generated by the SVD, and $\lambda$ is the singular value spectrum. As a result, the density operator can be transformed into a three-dimensional brick-wall tensor network (TN) \cite{ODTNS}, $ \rho(\beta) = Tr_P(\{Tr_G [\prod_{i \in \mathbb{A}} (G^{L(i)}_{p_ip_i',g(i)}) \prod_{j \in \mathbb{B}} (G^{R(j)}_{p_jp_j',g(j)}) ]\}^{3K}) $, where $G^{L(i)} _{p_ip_i',g(i)} = U^{L}_{p_ip_i',g} \sqrt{\lambda_{g}}$, $G^{R(j)}_{p_jp_j',g(j)}=U^R_{p_jp_j',g} \sqrt{\lambda_{g}}$, $\mathbb{A}$ and $\mathbb{B}$ stand for the two sublattices of the honeycomb lattice, and $Tr_{P(G)}$ means the trace over all shared physical (geometrical) bonds. Translational invariance is applied.

\begin{figure}[tbp]
\includegraphics[angle=0,width=1\linewidth]{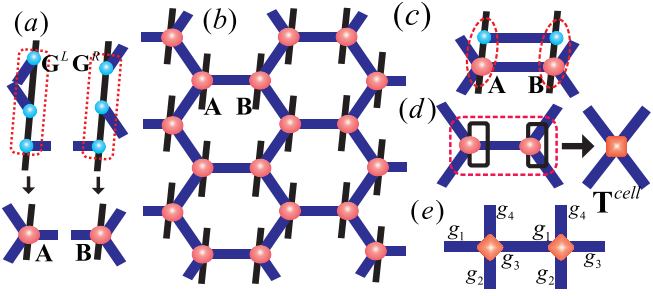}
\caption{(Color online) (a) By contracting three $\mathbf{G}^{L}$'s and $\mathbf{G}^{L}$'s together, respectively, we obtain the two inequivalent tensors $\mathbf{A}$ and $\mathbf{B}$ of the TPDO as shown in (b). (c) The TPDO is evolved by contracting a pair of $\mathbf{G}^L$ and $\mathbf{G}^R$ to the tensors $\mathbf{A}$ and $\mathbf{B}$. (d) After tracing over the physical bonds, we obtain the cell tensor $\mathbf{T}^{cell}$ by contracting a shared bond of $\mathbf{A}$ and $\mathbf{B}$. (e) The bond $g_3$ of the left $\mathbf{T}^{cell}$ connects the bond $g_1$ of the right $\mathbf{T}^{cell}$, thus $g_1$ and $g_3$ form an index pair.}
\label{fig-TPDO}
\end{figure}

To make the contractions, we represent the density operator at infinitesimal inverse temperature $\tau$ in the form of TPDO as $\rho(\tau) = Tr_G(\prod_{i\in \mathbb{A}} \mathbf{A}^i \prod_{j\in \mathbb{B}} \mathbf{B}^j)$ with $A^{i}_{p_ip_i',g_1g_2g_3} = \sum_{p_i''p_i'''} G^{L}_{p_ip_i'',g_1} G^{L}_{p_i''p_i''',g_2} G^{L}_{p_i'''p_i',g_3}$ and $B^{j}_{p_jp_j',g_1g_2g_3} = \sum_{p_j''p_j'''} G^{R}_{p_jp_j'',g_1} G^{R}_{p_j''p_j''',g_2} G^{R}_{p_j'''p_j',g_3}$ [Figs. \ref{fig-TPDO} (a) and (b)]. Hereafter, we use a capital letter in bold (with or without a superscript that is just a symbol to distinguish different tensors, e.g. $\mathbf{A}^i$) to denote a tensor \cite{tensor} and use the same capital letter with subscripts (which represent the tensor's bonds, e.g. $A^{i}_{p_ip_i',g_1g_2g_3}$) to denote the tensor element-wise.

To obtain $\rho(\beta)$ from $\rho(\tau)$, three pairs of $\mathbf{G}^L$ and $\mathbf{G}^R$ should be contracted repeatedly with the tensors $\mathbf{A}$ and $\mathbf{B}$ for the imaginary time evolution. We take the contraction of the pair in the $g_1$ direction as an example [Fig. \ref{fig-TPDO} (c)], which gives
\begin{eqnarray}
 A'_{pp', \tilde{g}_1g_2g_3} = \sum_{p''} A_{pp'',g_1g_2g_3} G^{L}_{p_i''p_i',g_1'}, \nonumber \\
 B'_{pp', \tilde{g}_1g_2g_3} = \sum_{p''} B_{pp'',g_1g_2g_3} G^{R}_{p_i''p_i',g_1'},
\label{eq-Evolution}
\end{eqnarray}
where $\tilde{g}_1=(g_1,g_1')$ is a composite bond. The contractions on other bonds of $\mathbf{A}$ and those of $\mathbf{B}$ are similar. During the contraction, the dimensions of geometrical bonds will be enlarged exponentially, and thus, proper truncations are needed to bound the dimensions. With obtaining $\hat{\rho}(\beta)$, the partition function $Z(\beta)=Tr[\hat{\rho}(\beta)]$ becomes the contraction of the infinite 2D TN formed by $\mathcal{A}_{g_1g_2g_3} = \sum_{p} A_{pp,g_1g_2g_3}$ and $\mathcal{B}_{g_1g_2g_3} = \sum_{p} B_{pp,g_1g_2g_3}$, namely
\begin{eqnarray}
 Z = Tr_G(\prod_{i\in \mathbb{A}} \mathbf{\mathcal{A}}^i \prod_{j\in \mathbb{B}} \mathbf{\mathcal{B}}^j).
\label{eq-CTN}
\end{eqnarray}
Similarly, the thermal average $\langle \hat{O}(\beta) \rangle = Tr[ \hat{\rho}(\beta) \hat{O} ]/Z(\beta)$ can also be calculated in the same way.

\section{Network contractor dynamics}

We now present the theory of NCD that simplifies the infinite contraction of the TN [Eq. (\ref{eq-CTN})] by a local contraction with a tensor cluster and a set of contractors. To proceed, we can transform the 2D TN so that it consists of only one inequivalent tensor $\mathbf{T}^{cell}$, and then introduce a set of unit-norm vectors which satisfy the fixed point equations of the mapping defined by $\mathbf{T}^{cell}$. The vectors at the fixed point are dubbed as \textit{contractors}. By substituting the TN with a tree-like defective TN, we obtain $Z(\beta)$ in terms of a simple local contraction of $\mathbf{T}^{cell}$ and the contractors.

\begin{figure}[tbp]
\includegraphics[angle=0,width=0.85\linewidth]{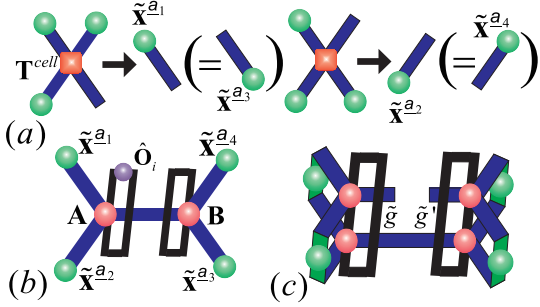}
\caption{(Color online) (a) The contractors satisfy the fixed point condition [Eq. (\ref{eq-FixedPoint})]. Due to the permutation invariance, when $i$ and $j$ belong to an index pair, the contractors satisfy $\mathbf{x}^{\alpha_i} = \mathbf{x}^{\alpha_j}$. (b) The thermal average of an operator $\hat{O}_i$ [Eq. (\ref{eq-Oi})]. (c) The environment matrix of the enlarged bond $\tilde{g}$ [Eq. (\ref{eq-SimplifiedM})].}
\label{fig-FixedPoint}
\end{figure}

By contracting the two inequivalent tensors into one tensor $\mathbf{T}^{cell}$ in the TN [Fig. \ref{fig-TPDO} (d)], one can get
\begin{eqnarray}
 T^{cell}_{g_1 g_2 g_3 g_4} = \sum_{g_5} \mathcal{A}_{g_1 g_2 g_5} \mathcal{B}_{g_3 g_4 g_5}.
\label{eq-CellT1}
\end{eqnarray}
In this case, there only exists one inequivalent tensor in $Z = Tr_{G} (... T^{cell}_{g_1 g_2 g_3 g_4} T^{cell}_{g_4 g_5 g_6 g_7} ...)$, which is a square TN. The $\mathbf{T}^{cell}$ has an important property that is indispensable to the simplifications of the TN contraction in the NCD. When the $i$th bond of $\mathbf{T}^{cell}$ is shared with the $j$th bond of its adjacent $\mathbf{T}^{cell}$, $\mathbf{T}^{cell}$ is invariant under the permutation of $i$th and $j$th two bonds. To take the above $\mathbf{T}^{cell}$ as an example, we have $T^{cell}_{g_1 g_2 g_3 g_4} = T^{cell}_{g_3 g_2 g_1 g_4} = T^{cell}_{g_1 g_4 g_3 g_2}$, where $g_1$ and $g_3$ ($g_2$ and $g_4$) form a pair [Fig. \ref{fig-TPDO} (e)]. We call such bonds $g_i$ and $g_j$ an \textit{index pair}. Now we introduce a vector set $\{ \mathbf{x}^{\alpha_{i}} \}$ containing $\mathcal{D}$ unit-norm vectors defined in the geometrical bond space of $\mathbf{T}^{cell}$ with $\mathcal{D}$ the order of $\mathbf{T}^{cell}$. Each $\mathbf{x}^{\alpha_{i}}$ has the same dimension with the $i$th bond of $\mathbf{T}^{cell}$. A set of mappings $\mathcal{T}_i$ ($i=1,...,\mathcal{D}$): $\prod_{\bigotimes a \neq i} \mathbb{V}^a \rightarrow \mathbb{V}^j$ is denoted as $\mathcal{T}_i (\{\mathbf{x}^{\alpha_{a \neq i}}\}) =\Gamma \mathbf{x}'^{\alpha_j}$ and defined by $\mathbf{T}^{cell}$ as
\begin{eqnarray}
\sum_{\{ g_{a \neq i} \}} T^{cell}_{g_1g_2g_3g_4} \prod_{\bigotimes a \neq i} x^{\alpha_a}_{g_{a}} = \Gamma x'^{\alpha_j}_{g_i},
\label{eq-Mapping}
\end{eqnarray}
where $\mathbb{V}^a$ denotes the space of $\mathbf{x}^{\alpha_a}$, $\Gamma$ is a positive real number to ensure that $\mathbf{x}'^{\alpha_j}$ is a unit-norm vector and the bonds $i$ and $j$ form an index pair. In the above definition, $(\mathcal{D}-1)$ vectors are contracted with $\mathbf{T}^{cell}$ except for $\mathbf{x}^{\alpha_i}$ and this contraction results in $\mathbf{x}'^{\alpha_j}$, a vector on bond $j$, while the bond $j$ forms a pair with bond $i$ [Fig. \ref{fig-FixedPoint} (a)]. For example, the mapping $\mathcal{T}_1 (\{\mathbf{x}^{\alpha_{a \neq 1}}\}) =\Gamma \mathbf{x}'^{\alpha_3}$ can be written as $\sum_{g_2g_3g_4} T^{cell}_{g_1g_2g_3g_4} x^{\alpha_2}_{g_2} x^{\alpha_3}_{g_3} x^{\alpha_4}_{g_4} = \Gamma x'^{\alpha_3}_{g_1}$, element-wise. We can write these $\mathcal{D}$ mappings in a more compact form as $ \{ \mathbf{x}' \} = \mathbf{\mathcal{T}}( \{ \mathbf{x} \})$, which means acting each $\mathbf{\mathcal{T}}^i$ once on the $ \{ \mathbf{x} \}$ to renew the $\mathcal{D}$ vectors.

\begin{figure}[tbp]
\includegraphics[angle=0,width=1\linewidth]{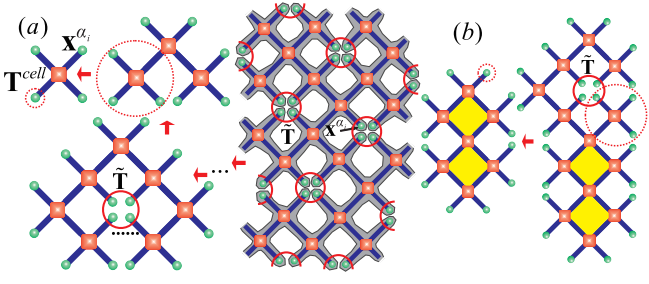}
\caption{(Color online) (a) The infinite defective TN of the partition function $Z$ can be contracted as a tree TN, because the $\tilde{\mathbf{T}}$'s (the rank-1 approximation of $\mathbf{T}^{cell}$ formed by the direct product of contractors) break the loops of the original TN, as shown by the shaded area. By repeatedly using the fixed point conditions [Eq. (\ref{eq-FixedPoint})], this contraction leads to Eq. (\ref{eq-PartitionZ}), giving a local contraction of $\mathbf{T}^{cell}$ and contractors. (b) Another possible construction of the defective TN with a tensor cluster that contains loops (yellow areas) of $\mathbf{T}^{cell(1)}$'s.}
\label{fig-ProveM}
\end{figure}

If $\mathbf{\mathcal{T}}$ maps a set of vectors $\{ \mathbf{\tilde{x}} \}$ into themselves [Fig. \ref{fig-TPDO} (e)] such that
\begin{eqnarray}
\mathbf{\mathcal{T}}(\{ \mathbf{\tilde{x}} \})=\tilde{\Gamma} \{ \mathbf{\tilde{x}} \},
\label{eq-FixedPoint}
\end{eqnarray}
then $\{ \mathbf{\tilde{x}} \}$ is the fixed point of $\mathbf{\mathcal{T}}$, where $\tilde{\Gamma} = Tr_G( \prod_{\bigotimes a} \tilde{\mathbf{x}} ^{\alpha_a} \mathbf{T}^{cell})$ is a positive real number to keep the vectors in $\{ \mathbf{\tilde{x}} \}$ with unit norm. We dub so-defined $\{ \mathbf{\tilde{x}} \}$ as the \textit{contractors} of the TN. Importantly, $\tilde{\mathbf{x}} ^{\alpha_i} = \tilde{\mathbf{x}} ^{\alpha_j}$ when bonds $i$ and $j$ form a pair due to the symmetry of the tensor. Consequently, we have the fixed point conditions in another form as $\mathcal{T}_i (\{\mathbf{x}^{\alpha_{a \neq i}}\}) =\Gamma \mathbf{x}^{\alpha_i}$ for $i=1,...,\mathcal{D}$, which are just the conditions for the rank-1 decomposition of $\mathbf{T}^{cell}$, i.e. the rank-1 tensor
\begin{eqnarray}
\tilde{\mathbf{T}} = \prod_{\bigotimes \alpha_a} \tilde{\mathbf{x}}^{\alpha_a}
\label{eq-Rank1T}
\end{eqnarray}
is the solution of $\min_{rank(\tilde{T})=1} | \mathbf{T}^{cell} - \tilde{\mathbf{T}} |$, where $|\bullet|$ stands for the norm of a tensor \cite{Rank1}. With the help of $\tilde{\mathbf{T}}$, we can approximate $Z$ by replacing the minimal number of $\mathbf{T}^{cell}$'s with $\tilde{\mathbf{T}}$'s (called \textit{defects}) so that there are no loops formed only by $\mathbf{T}^{cell}$'s. One can see in Fig. \ref{fig-ProveM} (a) that, the area marked by the grey shadow has no loop, so the contraction of this ``defective'' TN can be done as easily as that of a tree TN.

What's more, instead of actually carrying out the infinite contraction of the defective TN for calculating $Z$, one only needs to do a local contraction of $\mathbf{T}^{cell}$ and $\mathbf{\tilde{x}}$ thanks to the fixed point condition [Eq. (\ref{eq-FixedPoint})] as
\begin{eqnarray}
 Z \simeq \tilde{\Gamma}^{\mathcal{N}-1} Tr_{G}(\mathbf{T}^{cell} \tilde{\mathbf{T}}) = \tilde{\Gamma}^{\mathcal{N}-1} Tr_{G}(\mathbf{T}^{cell} \prod_{\bigotimes \alpha_a} \tilde{\mathbf{x}}^{\alpha_a}) = \tilde{\Gamma}^{\mathcal{N}},
\label{eq-PartitionZ}
\end{eqnarray}
with $\mathcal{N}$ the number of $\mathbf{T}^{cell}$'s [Fig. \ref{fig-ProveM} (a)]. This simplification can also be seen more clearly by reversing the contraction to a growing procedure that reconstructs the infinite defective TN in the following way: starting from Eq. (\ref{eq-PartitionZ}), Eq. (\ref{eq-FixedPoint}) is employed to replace one contractor with a $\mathbf{T}^{cell}$ and three contractors repeatedly (dash circles in Fig. \ref{fig-ProveM}). During the growing procedure, some contractors are involved to construct the rank-1 tensor $\tilde{\mathbf{T}}$ [Eq. (\ref{eq-Rank1T})], and by doing so, the defective TN is finally reconstructed with minimal numbers of defects. This growing picture also indicates that the choice of the replacement (approximating $\mathbf{T}^{cell}$'s by $\tilde{\mathbf{T}}$'s) is not unique, and as long as all loops of $\mathbf{T}^{cell}$'s are destroyed, Eq. (\ref{eq-PartitionZ}) holds. It should be remarked that the rank-1 tensor that appears in the defective TN is $T'_{g_1 g_2 g_3 g_4}=x^{\alpha_3}_{g_1} x^{\alpha_4}_{g_2} x^{\alpha_1}_{g_3} x^{\alpha_2}_{g_4}$, which equals to $\tilde{T}_{g_1 g_2 g_3 g_4}$ with permutation invariance in each index pair.

The thermal average of an operator $\hat{O}$ can be similarly calculated within the NCD scheme. It is noticed that, except for $\mathbf{T}^{cell}$'s that share the physical space with $\hat{O}$, the rest of the TN are exactly the same as those of the partition function $Z$. Consequently, $Tr(\hat{O}\hat{\rho})$ is the contraction of a tensor cluster formed by $\hat{O}$, related tensors and the contractors on the boundary. For instance, $\langle \hat{O}_{i} \rangle$ can be obtained [Fig. \ref{fig-FixedPoint} (b)] by
\begin{eqnarray}
  \langle \hat{O}_{i} \rangle =  \sum_{\scriptstyle p_ip_i'p_j \atop \scriptstyle g_1g_2g_3g_4g_5} A_{p_ip_i',g_1g_2g_5} B_{p_jp_j,g_3g_4g_5}  O_{p_ip_i'} x^{\alpha_1}_{g_1} x^{\alpha_2}_{g_2} x^{\alpha_3}_{g_3} x^{\alpha_4}_{g_4} /Z,
\label{eq-Oi}
\end{eqnarray}
where $O_{p_ip_i'} = \langle p_i| \hat{O}_{i} |p_i' \rangle$.

To describe the convergence of the nonlinear mapping and the stability of the fixed point, the Lyapunov exponent $I^{lya}$ can be introduced by
\begin{eqnarray}
 I^{lya} = \lim_{\Theta\rightarrow \infty} \sum_{\theta=1}^{\Theta} \{ \ln \sum_{a=1}^{\mathcal{D}} \cfrac {|\mathcal{T}^{\theta}(\{ \mathbf{x}_0 \}) - \mathcal{T}[ \mathcal{T}^{\theta-1}(\{ \mathbf{x}_0 \})+ \mbox{\boldmath$\epsilon$}^{a} ]|} {\mathcal{D} |\mbox{\boldmath$\epsilon$}^{a}|} \}/\Theta,
\label{eq-Lyapunov}
\end{eqnarray}
where $\mbox{\boldmath$\epsilon$}^{a}$ is the infinitesimal random vector to exert a perturbation on $\mathbf{x}^{\alpha_a}$ and $\mathcal{T}^{\theta} = \mathcal{T}[\mathcal{T}(\ldots )]$. The choice of $\{ \mathbf{x}_0 \}$ can be arbitrary when the total mapping time $\Theta$ is sufficiently large. The smaller $I^{lya}$, the faster $\{\mathbf{x}\}$ can through the mappings approach the fixed point against the perturbations.

As the defective TN in the above scheme is grown from a single $\mathbf{T}^{cell}$ and the contractors [Fig. \ref{fig-ProveM} (a)], another route for the construction of the defective TN is possible. Specifically speaking, one may begin with a cluster of $\mathbf{T}^{cell}$'s and contractors on its boundary. This cluster is allowed to contain loops of $\mathbf{T}^{cell}$, e.g. the cluster shown in Fig. \ref{fig-ProveM} (b). Notice that except the starting cluster, the rest of the TN will not contain any loops of $\mathbf{T}^{cell}$'s. This is relatively more accurate than the standard defective TN approximation, since one needs to replace certain $\tilde{\mathbf{T}}$'s in the standard defective TN back with $\mathbf{T}^{cell}$'s to recover corresponding loops. Thus, more loops recovered, more missing terms retrieved, and consequently, a higher accuracy can be achieved. However, the computational cost to contract such a cluster increases exponentially with the number of the loops inside, so in the following we shall propose an available scheme to recover the infinite loops. By doing so, Eq. (\ref{eq-Oi}) should be modified correspondingly.

Up to now we have established the NCD scheme upon the only requirement that the TN is represented in the form that consists of one inequivalent tensor which obeys permutation invariance in each index pair. For other choices of the inequivalent tensor of the prototype TN or for other TN's with representations that satisfy such requirement, the present NCD theory can apply straightforwardly.

\section{Cell tensor size, quasi-entanglement entropy and loop character}

\begin{figure}[tbp]
\includegraphics[angle=0,width=0.9\linewidth]{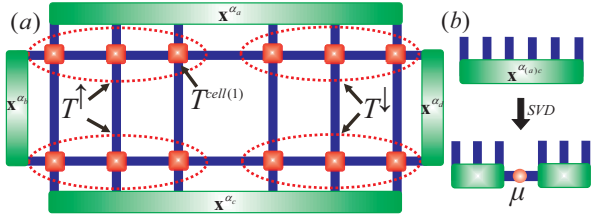}
\caption{(Color online) (a) A sketch of increasing the size of the cell tensor. Three $\mathbf{T}^{cell(1)}$'s are contracted to get $\mathbf{T}^{\uparrow }$ and $\mathbf{T}^{\downarrow }$ (dash circles), and then by Eq. (\ref{eq-IncreaseSize}), the whole cluster is considered as a cell tensor with $\gamma=12$. The green blocks with blue bonds represent corresponding contractors. (b) From the SVD of the contractor $\mathbf{x} ^{\alpha_{a(c)}, \gamma}$, we have the singular value spectrum $\mu$, and the QEE can be obtained with $\mu$ through Eq. (\ref{eq-Ent}).}
\label{fig-TcellSize}
\end{figure}

In the preceding section, we introduced the NCD theory using a cell tensor constructed as Eq. (\ref{eq-CellT1}), which contains the minimal number of original tensors in the TN. In this section, considering that the TN is only required in the form that consists of one inequivalent tensor, and recovering loops can improve accuracy, we may make different choices for constructing larger cell tensors in the NCD. We use $\mathbf{T}^{cell(\gamma)}$ to denote the cell tensor containing $\gamma$ cell tensors of the smallest size (denoted by $\mathbf{T}^{cell(1)}$), where $\gamma$ is called the cell tensor size. The NCD scheme can be directly used with $\mathbf{T}^{cell(\gamma)}$ as it gives no restriction for the choice of the cell tensor. Meanwhile, the defective TN obtained with $\mathbf{T}^{cell(\gamma)}$ ($\gamma>1$) has no loops of $\mathbf{T}^{cell(\gamma)}$ but it indeed contains loops of $\mathbf{T}^{cell(\gamma'<\gamma)}$. This suggests that the approximation becomes more accurate by choosing a cell tensor with larger size. To increase $\gamma$, we may construct $\mathbf{T}^{cell}$ through $\mathbf{T}^{\uparrow}$ and $\mathbf{T}^{\downarrow}$ as
\begin{eqnarray}
 T^{cell(\gamma)}_{(a_1a_3)(b_3b_4)(c_2c_4)(d_1d_2)} &=& \sum_{a_2a_4b_1b_2c_1c_3d_3d_4} T^{\uparrow \gamma_1}_{a_1b_1c_1d_1} T^{\uparrow \gamma_1}_{a_2b_2c_2d_2} T^{\downarrow \gamma_2}_{a_3b_3c_3d_3} \nonumber \\&& T^{\downarrow \gamma_2}_{a_4b_4c_4d_4} \delta_{a_2c_1} \delta_{b_1d_3} \delta_{a_4c_3} \delta_{b_2d_4},
\label{eq-IncreaseSize}
\end{eqnarray}
with $\gamma = 2(\gamma_1+\gamma_2)$. $\mathbf{T}^{\uparrow }$ and $\mathbf{T}^{\downarrow }$ are initiated as $\mathbf{T}^{cell(1)}$ and increased as $T^{\uparrow (\gamma+1)}_{(aa')b'(cc')d} = \sum_{bd'} T^{\uparrow (\gamma)}_{abcd} T^{cell(1)}_{a'b'c'd'} \delta_{bd'}$ and $T^{\downarrow (\gamma+1)}_{(aa')b(cc')d'} = \sum_{b'd} T^{\downarrow (\gamma)}_{abcd} T^{cell(1)}_{a'b'c'd'} \delta_{b'd}$, respectively, as shown in Fig. \ref{fig-TcellSize} (a). As is seen, the dimension of $T^{cell(\gamma)}$ increases exponentially with $\gamma$. Thus, truncations are needed in practical calculations, which are with the same principle for the truncations bounding the dimension during the contraction along the imaginary time [Eq. (\ref{eq-Evolution})]. The truncation principle is introduced in the following section.

Furthermore, by using Eq. (\ref{eq-IncreaseSize}), we introduce the ``transfer matrix" of the partition function, $\mathcal{M}^{(\gamma)} _{(a_1a_3) (c_2c_4)} = \sum_{b_3b_4 d_1d_2} T^{cell(\gamma)}_{(a_1a_3)(b_3b_4)(c_2c_4)(d_1d_2)} x^{\alpha_b,\gamma} _{b_3b_4} x^{\alpha_d,\gamma} _{d_1d_2}$, where two contractors $\mathbf{x} ^{\alpha_b, \gamma}$ and $\mathbf{x}^{\alpha_d,\gamma}$ are contracted with $\mathbf{T}^{cell(\gamma)}$. When $\gamma$ is sufficiently large, we can get $Z \simeq \lim_{l\rightarrow \infty} \tilde{\Gamma}(\gamma)^l Tr[\mathcal{M}(\gamma)^{l}]$ with $\tilde{\Gamma}(\gamma)$ a positive real number. Meanwhile, by Eq. (\ref{eq-FixedPoint}), we obtain $\tilde{\Gamma}(\gamma) x^{\alpha_a,\gamma} _{a_1a_3} = \sum_{c_2c_4} \mathcal{M} ^{(\gamma)} _{(a_1a_3) (c_2c_4)} x^{\alpha_c,\gamma} _{c_2c_4}$ and $\tilde{\Gamma}(\gamma) x ^{\alpha_c,\gamma} _{c_2c_4} = \sum_{a_1a_3} \mathcal{M} ^{(\gamma)}_{(a_1a_3) (c_2c_4)} x^{\alpha_a,\gamma} _{a_1a_3}$. This indicates that $\mathbf{x} ^{\alpha_{a(c)}, \gamma}$ is the dominant eigenstate of $\mathcal{M}^{(\gamma)}$ (remember $\alpha_a = \alpha_c$ due to the index pair rule). On the other hand, $\mathbf{x} ^{\alpha_{a(c)}, \gamma}$ can be written in the form of a matrix product state (MPS) \cite{MPS} with ($\gamma_1 +\gamma_2$) ``physical" bonds, and we may introduce the entanglement entropy of $\mathbf{x} ^{\alpha_{a(c)}, \gamma}$ that is closely related to the ``separability'' \cite{Separable} of states by
\begin{eqnarray}
S(\gamma) =-\sum_i(\mu_i^{a(c),\gamma})^2 \ln [(\mu_i^{a(c),\gamma})^2],
\label{eq-Ent}
\end{eqnarray}
where $\mu_i^{a(c),\gamma}$ is the normalized singular spectrum value of matrix $x ^{\alpha_{a},\gamma} _{a_1a_3}$ ($x ^{\alpha_{c},\gamma} _{c_2c_4}$) [Fig. \ref{fig-TcellSize} (b)]. We regard $S$, which is the entanglement entropy of the MPS representing the dominant eigenstate of the transfer matrix of $Z$, as the quasi-entanglement entropy (QEE) of the density operator.

To monitor the effects of loops of the $\mathbf{T}^{cell(\gamma)}$'s, we may introduce the loop character $I^{loop}(\gamma)$ defined as
\begin{eqnarray}
 I^{loop}(\gamma) =1-\sum_{bb'} x^{\alpha_b,\gamma+2}_{bb'} x^{\alpha_{b},\gamma}_{bb'},
\label{eq-LoopInfor}
\end{eqnarray}
where $\mathbf{x}^{\alpha_i,\gamma}$ are the contractors of $\mathbf{T}^{cell(\gamma)}$. This quantity describes the difference between the contractors of the cell tensors of different sizes, that are $(\gamma+2)$ and $\gamma$. It is worth mentioning that while $I^{lya}$ delineates the properties of a certain mapping, $I^{loop}$ describes the difference between two mappings with different sizes of the cell tensor. Specifically, when $I^{loop}(\gamma)$ converges to zero as $\gamma$ increases, it means the cell tensor converges (i.e. $\mathbf{T}^{cell(\gamma)} = \mathbf{T}^{cell(\gamma'>\gamma)}$), and thus a further increase of $\gamma$ will not change the results, which are obtained by the contraction of the cell tensor and the contractors [Eqs. (\ref{eq-PartitionZ}) and (\ref{eq-Oi})]. Remember that the defective TN of $\mathbf{T}^{cell(\gamma+2)}$ contains loops of $\mathbf{T}^{cell(\gamma)}$, while the defective TN of $\mathbf{T}^{cell(\gamma)}$ does not, thus the effects from loops larger than $\gamma$ are negligible and the present defective TN is a good approximation of the original one. Therefore, $I^{loop}(\gamma)$ can be used to monitor the error brought by the defective TN approximation, or in other words, the error brought by replacing $\mathbf{T}^{cell(\gamma)}$'s with the rank-$1$ approximations.


\section{Implementation of NCD scheme}

For the implementation of the NCD, we propose the imaginary-time-sweep algorithm (ITSA) where the sweep procedure along the imaginary time ensures that the truncations are optimal, i.e. they are obtained in consideration of the whole TN of $\hat{\rho} (\tilde{\beta})$ at the targeted temperature $\tilde{\beta}$. To sweep, we construct a double layer TN for $Z(\tilde{\beta})$ with $Z(\tilde{\beta}) = Tr[\hat{\rho} (\tilde{\beta})] = Tr[\hat{\rho} (\beta) \hat{\rho} (\tilde{\beta}-\beta)]$. The cell tensor is obtained as $T^{cell}_{\xi_1 \xi_2 \xi_3 \xi_4} = \sum_{\xi_5} \mathcal{A}_{\xi_1 \xi_2 \xi_5} \mathcal{B}_{\xi_3 \xi_4 \xi_5}$, where $\mathcal{A}_{\xi_1 \xi_2 \xi_5} = \sum_{pp'} A_{pp',g_1g_2g_5} \tilde{A} _{p'p, \tilde{g}_1 \tilde{g}_2 \tilde{g}_5}$ and $\mathcal{B}_{\xi_3 \xi_4 \xi_5} = \sum_{pp'} B_{pp',g_3g_4g_5} \tilde{B} _{p'p, \tilde{g}_3 \tilde{g}_4 \tilde{g}_5}$. $\mathbf{A}$ and $\mathbf{B}$ ($\tilde{\mathbf{A}}$ and $\tilde{\mathbf{B}}$) are the two inequivalent tensors of $\hat{\rho} (\beta)$ [$\hat{\rho} (\tilde{\beta}-\beta)$] and $\xi_a = (g_a,\tilde{g}_a)$ is a composite bond. The double layer TPDO has the same form as that of the single layer, thus the NCD scheme is directly applicable.

Before we present the ITSA, we explain the truncation principle to limit the computational costs for practical calculations, and show how to make the optimal truncations with NCD. To obtain the optimal truncation of an enlarged bond $\tilde{g}$ with its dimension $\tilde{\chi}$ on the TPDO of $\hat{\rho} (\beta)$, we consider the matrix $\mathbf{M}^e$ such that $Z = \sum_{\tilde{g}\tilde{g}'} \delta_{\tilde{g}\tilde{g}'} M^e_{\tilde{g}\tilde{g}'} = Tr(\mathbf{M}^e)$, which is obtained by simply contracting all shared bonds in the TN of $Z$ except $\tilde{g}$. We dub $\mathbf{M}^e$ as the environment matrix of the enlarged bond $\tilde{g}$. Note that the idea of using the environment of a tensor or a tensor cluster to get a non-local optimal truncation is already well-known (e.g. Refs. [\onlinecite{PEPS}] and [\onlinecite{TRG3}]). According to the linear algebra, the best truncation of a matrix can be reached using the SVD, say $\mathbf{M}^e \simeq \mathbf{P} \mathbf{\Lambda} \mathbf{Q}^T$, where only $\chi$ largest singular values and the corresponding left and right singular vectors are kept. Here $\chi$ is the preset dimension cut-off. This truncation globally minimizes $|Z-Tr(\mathbf{P} \mathbf{\Lambda} \mathbf{Q}^T)|$. The truncation error can be controlled by
\begin{eqnarray}
 \varepsilon = (\sum_{a=\chi+1}^{\tilde{\chi}} \Lambda_{a}) / (\sum_{a=1}^{\tilde{\chi}} \Lambda_{a}).
\label{eq-TrunctionErr}
\end{eqnarray}
Redefining the matrix as $\mathbf{\tilde{M}} = \sqrt{\mathbf{\Lambda}} \mathbf{P}^T \mathbf{Q} \sqrt{\mathbf{\Lambda}}$, we have $Z \simeq Tr(\mathbf{\tilde{M}})$, and then use the SVD again to decompose $\mathbf{\tilde{M}}$ as $\mathbf{\tilde{M}} = \mathbf{\tilde{P}} \mathbf{\tilde{\Lambda}} \mathbf{\tilde{Q}}^T$. The optimal $\tilde{\chi} \times \chi$ truncation matrices (which are in fact isometries) to project the dimensions of bonds $\tilde{g}$ and $\tilde{g}'$ from $\tilde{\chi}$ to $\chi$ are obtained by
\begin{eqnarray}
 \mathbf{\check{P}} = \mathbf{P} \mathbf{\Lambda}^{-1/2} \mathbf{\tilde{P}} \sqrt{\mathbf{\tilde{\Lambda}}}, \
 \mathbf{\check{Q}} = \mathbf{Q} \mathbf{\Lambda}^{-1/2} \mathbf{\tilde{Q}} \sqrt{\mathbf{\tilde{\Lambda}}}.
\label{eq-TruncationMQ}
\end{eqnarray}
$\mathbf{\check{P}}$ and $\mathbf{\check{Q}}$ form an identical transformation as $\mathbf{\check{P}} \mathbf{\check{Q}}^T \simeq \mbox{\boldmath$\delta$}$, where $\mbox{\boldmath$\delta$}$ is the identity matrix.

Thus, as long as $\mathbf{M}^e$ is obtained, the truncation matrices can be reached. Obviously, the difficulty in obtaining $\mathbf{M}^e$ is as big as calculating $Z$ itself. However, in the NCD scheme with the help of Eq. (\ref{eq-FixedPoint}), the calculation of $\mathbf{M}^e$ can be remarkably simplified with the defective TN as
\begin{eqnarray}
 \cfrac {M^e_{g_jg'_j}} {\tilde{\Gamma}^{\mathcal{N}}}=
\left\{
\begin{array}{lll}
 \sum_{g''_j} x^{j}_{g_jg'_j} x^{j}_{g'_jg''_j} \\
 \sum_{\xi_{a}\xi_{b}\xi_{d}\xi_{f}g''_j} \mathcal{A}^i_{\xi_{a}\xi_{b} g_jg''_j} \mathcal{B}^j_{\xi_{d}\xi_{f} g'_jg''_j} x^{a}_{\xi_a} x^{b}_{\xi_b} x^{a}_{\xi_d} x^{b}_{\xi_f} / \tilde{\Gamma},
\end{array}
\right.
\label{eq-SimplifiedM}
\end{eqnarray}
where we write selectively the composite index $\xi_i$ into ($g_i,g'_i$) for clarity. The first line of the right-hand-side of Eq. (\ref{eq-SimplifiedM}) holds when the enlarged bond is one of $\mathbf{T}^{cell}$'s bonds, and the second line holds when the enlarged bond is contracted in the construction of $\mathbf{T}^{cell}$ [Fig. \ref{fig-FixedPoint} (c)]. The approximation of $\mathbf{M}^e$ can also be improved similarly by using a cluster like Fig. \ref{fig-ProveM} (b) or by increasing the cell tensor size.

\begin{figure}[tbp]
\includegraphics[angle=0,width=1\linewidth]{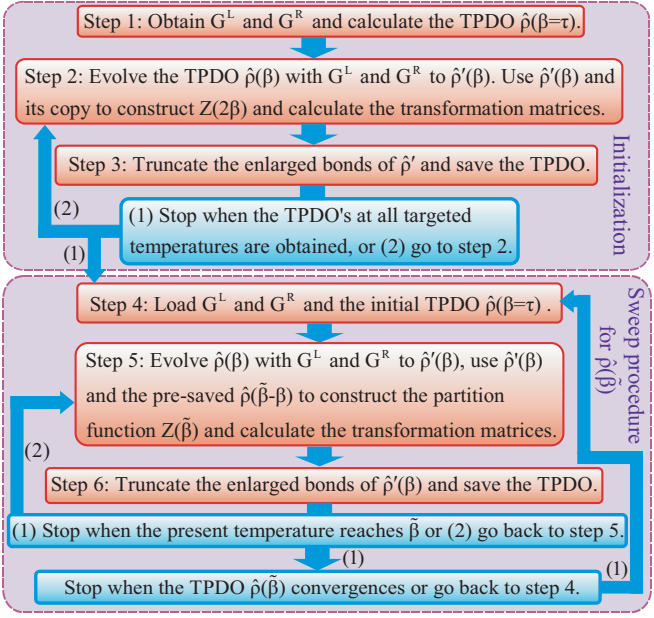}
\caption{(Color online) The flowchart of the ITSA algorithm.}
\label{fig-ITSA}
\end{figure}

Now we propose the ITSA, which has three parts: initialization, sweep procedure (Fig. \ref{fig-ITSA}) and calculations of observables. In the initialization, we begin with the TPDO $\hat{\rho}(\tau)$ and evolve it until the targeted temperature $\tilde{\beta}$ is reached. During the evolution at $\beta$ ($\tau < \beta \leq \tilde{\beta}$), the contractors are obtained with the power algorithm of the rank-1 decomposition and the truncation matrices are calculated with the double layer TPDO $Tr[\hat{\rho}(\beta)^2]$ [Eq. (\ref{eq-SimplifiedM})]. After each time of truncation, the TPDO is saved. In the sweep procedure, we start the evolution with $\hat{\rho}(\tau)$ again, and obtain $\mathbf{M}^e$ for each truncation from the TN of $Z(\tilde{\beta})$, i.e. we construct $Z(\tilde{\beta})$ using $\hat{\rho}(\beta)$ and $\hat{\rho} (\tilde{\beta}-\beta)$, where $\hat{\rho}(\beta)$ was obtained from the evolution while $\hat{\rho} (\tilde{\beta}-\beta)$ was obtained in the initialization or in the last sweep procedure. After the truncation, the TPDO of $\hat{\rho}(\beta)$ is renewed. With renewing the TPDO's at all $\beta$, we start the sweep with $\hat{\rho}(\tau)$ again unless $\hat{\rho}(\tilde{\beta})$ converges. With obtaining of $\hat{\rho}(\beta)$, $Z(\tilde{\beta})$ and $\langle \hat{O}(\tilde{\beta}) \rangle$ can be readily calculated with Eqs. (\ref{eq-PartitionZ}) and (\ref{eq-Oi}).

The errors are from three parts that are well controlled by three factors separately in the ITSA: the error of Trotter-Suzuki decomposition controlled by $\tau$ (infinitesimal imaginary time slice, see Sec. II), the truncation error controlled by $\varepsilon$ [Eq. (\ref{eq-TrunctionErr})] and the defective TN approximation controlled by the loop character $I^{loop}$ (see the arguments in Sec. IV). Importantly, the above sweep procedure escapes from the error accumulation during the imaginary time evolution as the truncations are obtained by minimizing the error of partition function at the targeted temperature.

\section{Applications of the NCD}

\begin{figure}[tbp]
\includegraphics[angle=0,width=0.8\linewidth]{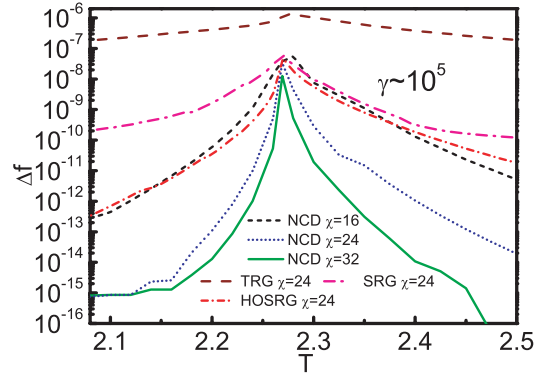}
\caption{(Color online) The temperature dependence of the relative error $\Delta f = |(f_{NCD} - f_{exact})/f_{exact}|$ of the free energy of the $2$D Ising model on square lattice calculated by the NCD ($\chi = 16, 24, 32$), TRG, SRG and HOSRG algorithms \cite{impHOSRG} ($\chi = 24$). We set the cell tensor size $\gamma \sim 10^{5}$ for the NCD calculations.}
\label{fig-Isingf}
\end{figure}

We first benchmark the efficiency and accuracy of the NCD scheme by using the exact result of the Ising model on square lattice. Fig. \ref{fig-Isingf} shows the relative error $\Delta f = |(f_{NCD} - f_{exact})/f_{exact}|$ of the free energy by the NCD scheme, tensor renormalization group (TRG),  second renormalization group (SRG) and higher-order second renormalization group (HOSRG) algorithms \cite{HOSRG,impHOSRG} against temperature for a comparison. One can see that the error of the NCD algorithm approaches rapidly the machine error (about $10^{-15}$) when temperature is away from the critical point, and reaches the maximum at the critical temperature, which is about $1.26 \times 10^{-8}$ for $\chi = 32$. The comparison shows that the accuracy of our NCD method is higher than that of the HOSRG away from the critical point, and in the vicinity of critical point the accuracies of NCD and HOSRG algorithms are comparable. As the accuracy of HOSRG is over the TRG, SRG and HOTRG methods,  which has already been discussed in Ref. [\onlinecite{HOSRG}], the accuracy of our method is over those methods particularly away from the critical point, as also manifested in Fig. \ref{fig-Isingf}.

\begin{figure}[tbp]
\includegraphics[angle=0,width=1\linewidth]{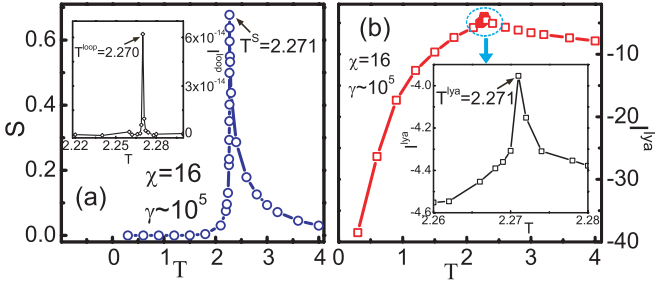}
\caption{(Color online) (a) The temperature dependence of the QEE $S$ of the 2D Ising model on square lattice, where the critical temperature $\textit{$T^S$} = 2.271$ is obtained. Inset: the temperature dependence of the loop character, showing $\textit{$T^{loop}$} = 2.270$. (b) The temperature dependence of the Lyapunov exponent of the $2$D Ising model, giving $\textit{$T^{lya}$} = 2.271$. The inset shows the detail near the critical point. The relative error of the critical temperature is about $10^{-3}$, where the exact critical temperature $T_c = 2/\ln (1+\sqrt{2}) \simeq 2.2692$.}
\label{fig-IsingEnt}
\end{figure}

It is also shown in Fig. \ref{fig-IsingEnt} that three quantities $S$, $I^{loop}$ and $I^{lya}$ proposed in the NCD scheme can be used to determine the critical temperature of the 2D Ising model by locating the maximum. The error of the critical temperature given by these three quantities is about $10^{-3}$, where the exact critical temperature $T_c = 2/\ln (1+\sqrt{2})$.

\begin{figure}[tbp]
\includegraphics[angle=0,width=1\linewidth]{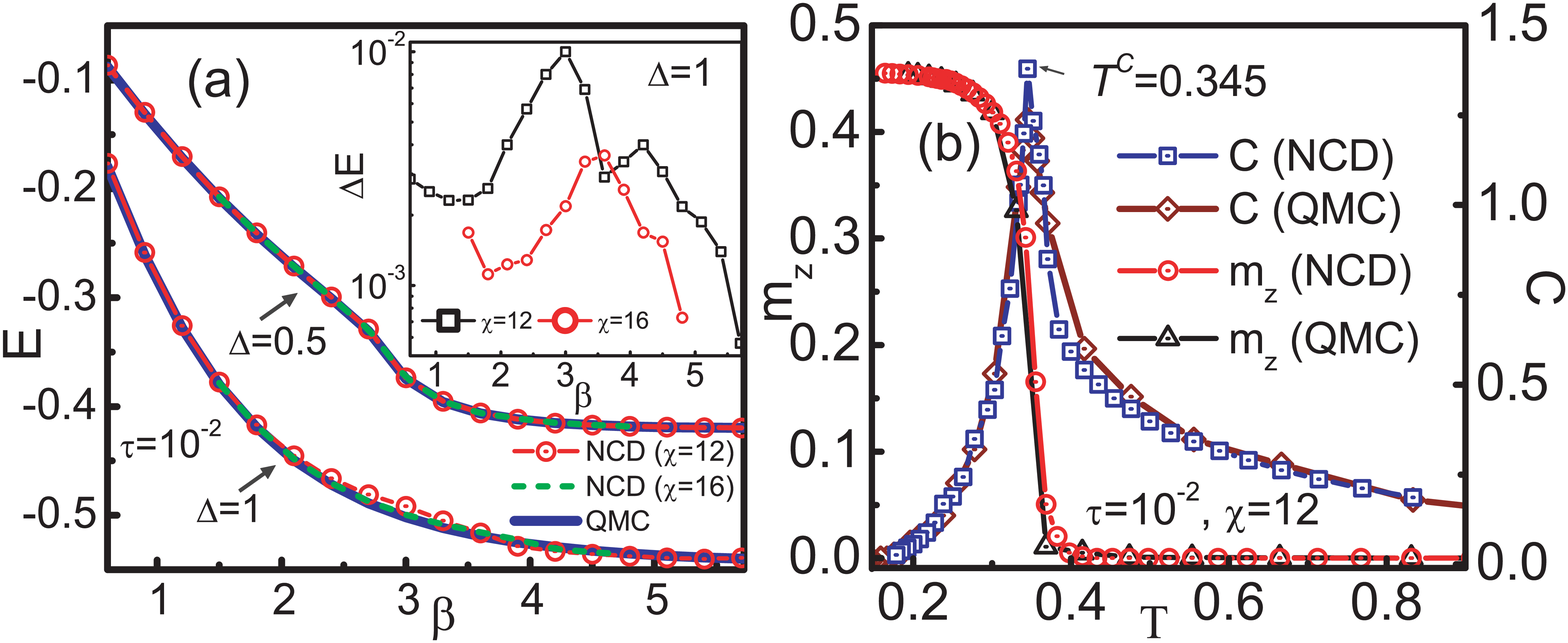}
\caption{(Color online) (a) Energy per site $E$ obtained by NCD and QMC of the spin-$1/2$ Heisenberg antiferromagnet on honeycomb lattice at $\Delta=0.5$ and $1$. Inset: the energy difference $\Delta E = |E_{NCD} - E_{QMC}|$ against temperature at $\chi=12$ and $\chi=16$. (b) The staggered magnetization per site $m_z$ and the specific heat $C$ as functions of temperature $T$ for $\Delta = 0.5$. The specific heat indicates a thermodynamic phase transition at $\textit{$T^C$} = 0.345$.}
\label{fig-Energy}
\end{figure}

To show the efficiency and accuracy of the NCD approach for calculating the thermodynamics of 2D quantum systems, we take the spin-1/2 HAF on honeycomb lattice with nearest neighbor interactions as an example, whose local Hamiltonian reads $\hat{H}^{ij} = \Delta [\hat{S}^{i(x)} \hat{S}^{j(x)} + \hat{S}^{i(y)} \hat{S}^{j(y)}] + \hat{S}^{i(z)} \hat{S}^{j(z)}$, where $\Delta$ characterizes the anisotropy of exchange interactions. First, we calculated the energy per site $E = Tr(\hat{H}_{ij} \hat{\rho})$ at $\Delta = 0.5$ and $1$ to testify the validity of NCD scheme. The results are well compared with QMC simulations, as shown in Fig. \ref{fig-Energy} (a), in which different dimension cut-offs $\chi$ are used in the NCD calculations. The inset shows that around the crossover point when there is no spin anisotropy ($\Delta = 1$) which is the most challenging parameter range \cite{ODTNS}, the energy difference $\Delta E$ betweens the results of the NCD and QMC is around $~ 10^{-3}$ for the dimension cut-off $\chi = 16$.

We used the tensor cluster in Fig. \ref{fig-ProveM} (b) to calculate the truncation matrices, where the truncation error $\varepsilon$ is found around $10^{-4} \sim10^{-6}$. In calculating the observables, the loop character $I^{loop}$ is required under $10^{-7}$ and the cell tensor size is increased until the difference between the observables of size $\gamma$ and $\gamma+2$ is less than $10^{-6}$. We use the power algorithm \cite{Rank1} to calculate the rank-$1$ decomposition and the iteration is stopped when the difference between the vectors at $t$ and $t+1$ steps is less than $10^{-14}$. For the characters of the TNS, we fix the cell tensor size $\gamma = 500$ and set $\Theta = 300$ for $I^{lya}$. We set the lattice size as $64 \times 64$ for QMC calculations and keep the errors around $10^{-5}$.

\begin{figure}[tbp]
\includegraphics[angle=0,width=1\linewidth]{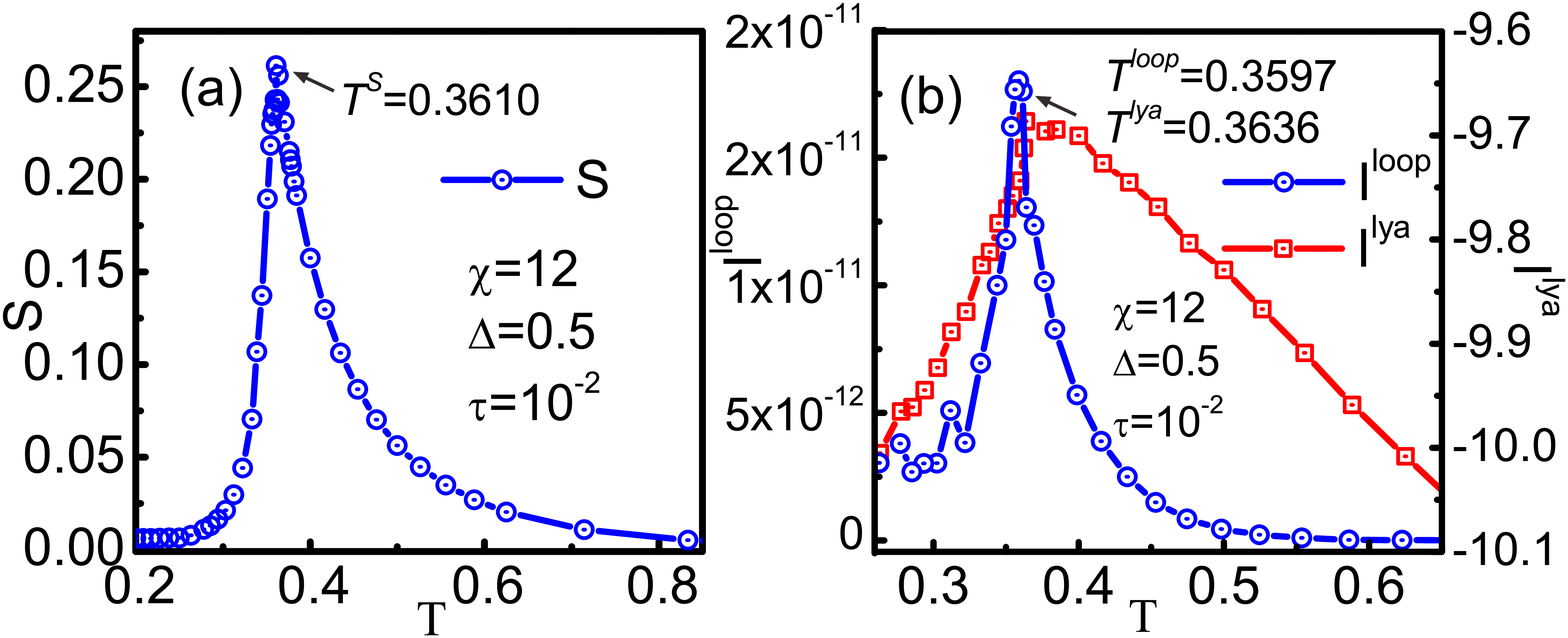}
\caption{(Color online) (a) The temperature dependence of the QEE S, where the critical temperature is found to be $\textit{$T^S$} = 0.3610$. (b) The temperature dependence of the Lyapunov exponent I$^{lya}$ and the loop character I$^{loop}$, which show sharp peaks at $\textit{$T^{lya}$} = 0.3636$ and $\textit{$T^{loop}$} = 0.3597$, respectively.}
\label{fig-TPT}
\end{figure}

It is known that in the present system a thermodynamic phase transition (TPT) may occur in the presence of anisotropy ($0\leq \Delta<1$), and the critical temperature can be determined by the divergent peak of specific heat $C = -\beta^2 d E/ d \beta$. Our calculation of $C$ shows that the TPT occurs at $\textit{$T^{C}$} = 0.345$ with $\Delta = 0.5$ [Fig. \ref{fig-Energy} (b)] for both NCD and QMC calculations. We also calculated the QEE, where the similar behavior near the critical point is observed [Fig. \ref{fig-TPT} (a)]. The sharp peak of $S$ appears at $\textit{$T^{S}$} = 0.3610$, close to the value obtained from the specific heat. The QEE vanishes (about $10^{-3} \thicksim 10^{-4}$), indicating that the dominant eigenstate of $Z$ becomes more separable \cite{Separable} when the temperature is away from the critical vicinity.

We studied the $T$-dependence of the Lyapunov exponent $I^{lya}$ and the loop character $I^{loop}$. The results [Fig. \ref{fig-TPT} (b)] show that the characters reach peaks at $\textit{$T^{lya}$} = 0.3636$ and $\textit{$T^{loop}$} = 0.3597$, again close to the critical temperatures obtained from the specific heat and the QEE.

We would like to draw some discussions about the QEE, $I^{lya}$ and $I^{loop}$, which all show peaks in the vicinity of the critical point for both classical and quantum systems. For the QEE, it is already known that the entanglement entropy of the MPS can detect phase transitions in one-dimensional quantum systems \cite{Ent} and $2D$ classical systems \cite{ZhaoY}. What we propose here is to detect phase transitions of $2D$ quantum systems using the similar idea. The difference is that the MPS is reached from the transfer matrix of $Z$ represented as a TN, which is obtained through the imaginary time evolution of the TPDO. The QEE describes the non-locality of the states, thus it is reasonable to find its maximum at the critical point.

The behaviors of $I^{lya}$ and $I^{loop}$ near in the critical vicinity are also related to the non-locality of the states. The value of $I^{loop}$ indicates how important the effects of larger loops are, as is argued in Sec. IV. From the results, we can see that the TPDO bears high loop dependence, i.e. the effects from the loops of cell tensors can be ignored only when the size of cell tensor $\gamma$ is large enough in the critical vicinity, which is coincident with the non-locality. $I^{lya}$ can be understood in the defective scheme. With a large $I^{lya}$, one can map any vectors close to the fixed point (the contractors $\{\tilde{\mathbf{x}}\}$) by acting the mapping $\mathcal{T}$ for great times [Eq. (\ref{eq-FixedPoint})]. One can see that one time of the mapping is equivalent to the contraction of one $\mathbf{T}^{cell}$ and the vectors for generating a new vector in the defective TN.  The larger the $I^{lya}$ is, the more time of the mapping is needed to reach the fixed point, which corresponds to the contraction of a larger defective TN. So the $I^{lya}$ is relatively large in the critical vicinity, which is also coincident with the non-locality of the state.

\section{Summary}

In summary, we developed the NCD theory for exploring the thermodynamic properties of $2$D quantum lattice models, and proposed the ITSA that is free from the negative sign problem for the numerical realization of NCD scheme. We benchmark the accuracy of the NCD scheme on the square Ising model and then calculate the thermodynamics of the spin-1/2 HAF on honeycomb lattice as an example, which are found consistent well with the exact results and the QMC simulations, respectively. New characters such as the quasi-entanglement entropy, Lyapunov exponent and loop character are introduced to describe properties of the thermal states and can be utilized to detect possible thermodynamic phase transitions of both classical and quantum systems. The straightforward extension of the NCD method to other classical or quantum lattice systems is possible.

\section*{Acknowledgements}
The authors are indebted to W. Li, X. Yan, Y. Zhao, Z. Y. Xie, H. H. Zhao, M. P. Qin, J. Chen, and T. Xiang for stimulating discussions. This work is supported in part by the NSFC (Grants No. 90922033 and No. 10934008), the MOST of China (Grant No. 2012CB932900 and No. 2013CB933401), and the CAS.


\begin{thebibliography}{99}
\bibitem{tensor} We generally consider vectors and matrices as the first- and second-order tensors.
\end{thebibliography}
\end{document}